\documentclass[dvips]{acta}
\usepackage{supertabular,lscape,epsfig}
\usepackage{amssymb}
\usepackage{amsmath}
\DeclareSymbolFont{ppa}{OT1}{ppl}{m}{it}
\DeclareMathSymbol{\vv}{\mathalpha}{ppa}{'166}

\newfont{\hb}{rphvb at 10pt}
\newfont{\hbo}{rphvbo at 10pt}
\newfont{\bitt}{rptmbi at 12pt}
\newfont{\bits}{rptmbi at 11pt}

\SetPages{91}{107}

\SetVol{60}{2010}

\begin{document}

\newcommand{\TabCapp}[2]{\begin{center}\parbox[t]{#1}{\centerline{
  \small {\spaceskip 2pt plus 1pt minus 1pt T a b l e}
  \refstepcounter{table}\thetable}
  \vskip2mm
  \centerline{\footnotesize #2}}
  \vskip3mm
\end{center}}

\newcommand{\TTabCap}[3]{\begin{center}\parbox[t]{#1}{\centerline{
  \small {\spaceskip 2pt plus 1pt minus 1pt T a b l e}
  \refstepcounter{table}\thetable}
  \vskip2mm
  \centerline{\footnotesize #2}
  \centerline{\footnotesize #3}}
  \vskip1mm
\end{center}}

\newcommand{\MakeTableSepp}[4]{\begin{table}[p]\TabCapp{#2}{#3}
  \begin{center} \TableFont \begin{tabular}{#1} #4 
  \end{tabular}\end{center}\end{table}}

\newcommand{\MakeTableee}[4]{\begin{table}[htb]\TabCapp{#2}{#3}
  \begin{center} \TableFont \begin{tabular}{#1} #4
  \end{tabular}\end{center}\end{table}}

\newcommand{\MakeTablee}[5]{\begin{table}[htb]\TTabCap{#2}{#3}{#4}
  \begin{center} \TableFont \begin{tabular}{#1} #5 
  \end{tabular}\end{center}\end{table}}

\newfont{\bb}{ptmbi8t at 12pt}
\newfont{\bbb}{cmbxti10}
\newfont{\bbbb}{cmbxti10 at 9pt}
\newcommand{\uprule}{\rule{0pt}{2.5ex}}
\newcommand{\douprule}{\rule[-2ex]{0pt}{4.5ex}}
\newcommand{\dorule}{\rule[-2ex]{0pt}{2ex}}
\def\thefootnote{\fnsymbol{footnote}}
\begin{Titlepage}
\Title{The Optical Gravitational Lensing Experiment.\\
The OGLE-III Catalog of Variable Stars.\\
VIII.~Type~II Cepheids in the Small Magellanic Cloud\footnote{Based on
observations obtained with the 1.3-m Warsaw telescope at the Las Campanas
Observatory of the Carnegie Institution of Washington.}}
\Author{I.~~S~o~s~z~y~ñ~s~k~i$^1$,~~
A.~~U~d~a~l~s~k~i$^1$,~~
M.\,K.~~S~z~y~m~a~ñ~s~k~i$^1$,\\
M.~~K~u~b~i~a~k$^1$,~~
G.~~P~i~e~t~r~z~y~ñ~s~k~i$^{1,2}$,~~
\L.~~W~y~r~z~y~k~o~w~s~k~i$^3$,\\
K.~~U~l~a~c~z~y~k$^1$~~
and~~ R.~~P~o~l~e~s~k~i$^1$}
{$^1$Warsaw University Observatory, Al.~Ujazdowskie~4, 00-478~Warszawa, Poland\\
e-mail:
(soszynsk,udalski,msz,mk,pietrzyn,kulaczyk,rpoleski)@astrouw.edu.pl\\
$^2$ Universidad de Concepción, Departamento de Astronomia, Casilla 160--C, Concepción, Chile\\
$^3$ Institute of Astronomy, University of Cambridge, Madingley Road, Cambridge CB3 0HA, UK\\
e-mail: wyrzykow@ast.cam.ac.uk}
\Received{May 21, 2010}
\end{Titlepage}
\Abstract{The eighth part of the OGLE-III Catalog of Variable Stars
(OIII-CVS) contains type~II Cepheids in the Small Magellanic Cloud (SMC).
The sample consists of 43 objects, including 17 BL~Her, 17~W~Vir and 9
RV~Tau stars (first examples ever found in the SMC). Seven stars have been
classified as peculiar W~Vir stars -- a recently identified subclass of
type~II Cepheids. These stars have distinctive light curves, are brighter
and bluer than the ordinary W~Vir variables. We confirm that a large
fraction of the peculiar W~Vir stars are members of binary systems.

Three type~II Cepheids exhibit eclipsing variations superimposed on the
pulsation light curves, and three other objects show long-period
ellipsoidal variability. All stars with the indication of binarity display
secondary periods which may be interpreted as amplitude and/or phase
modulations of the pulsation light curves with periods equal to the orbital
periods or half the orbital periods. We do not have any model for these
modulations, however this phenomenon rules out a possibility of the optical
blends of a pulsating star and a binary system.

For each object the multi-epoch {\it V}- and {\it I}-band photometry
collected over 8 or 13 years of observations and finding charts are
available to the astronomical community from the OGLE Internet archive.}
{Stars: variables: Cepheids -- Stars: oscillations --Stars: Population II
-- Magellanic Clouds}

\Section{Introduction}
Type~II Cepheids (also called Population II Cepheids) are low-mass
pulsating variable stars of the intermediate disk or halo population. They
are divided into three groups, each one at a different evolutionary stage:
BL~Her with the shortest periods, W~Vir with intermediate periods and
RV~Tau stars with the longest periods. BL~Her stars are believed to evolve
through the instability strip from the horizontal branch toward the
asymptotic giant branch (AGB, Strom \etal 1970). W~Vir stars loop into the
instability strip as a result of the helium-shell-flash episodes which
occur during the AGB stage (Schwarzschild and H{\"a}rm 1970, Gingold
1976). RV~Tau stars are post-AGB objects which cross the instability strip
during their rapid evolution toward the white dwarf stage (Gingold 1974,
Jura 1986).

To date, very few type~II Cepheids have been known in the Small Magellanic
Cloud (SMC). The first variable of this type was identified by Tifft (1963)
in the vicinity of the globular cluster NGC~121. The catalog of variable
stars in the SMC by Payne-Gaposchkin and Gaposchkin (1966) contained three
type~II Cepheids, of which one object -- HV~206 -- is now classified in the
General Catalogue of Variable Stars (GCVS, Artyukhina \etal 1995) as a
semiregular variable. Then, Ge{\ss}ner (1981) added XY~Hyi to the list of
known type~II Cepheids in the SMC. This modest sample of type~II Cepheids
in the SMC was extended by Udalski \etal (1999), who studied the data
collected during the second phase of the Optical Gravitational Lensing
Experiment (OGLE-II). The OGLE-II catalog of Cepheids in the SMC contained
a list of Cepheid-like variables which were fainter than fundamental-mode
classical Cepheids. In the present study we examine this list and 11
objects classify as type~II Cepheids. No RV Tau star in the SMC have been
reported so far.

In this paper we increase the number of known type~II Cepheids in the SMC
to 43. Our investigation is based on the photometric data collected during
the third phase of the OGLE project. This work is a part of the OGLE-III
Catalog of Variable Stars (OIII-CVS) which is intended to include all
variable sources monitored by the OGLE-III survey in the Magellanic Clouds
and Galactic bulge. In the previous papers of this series we presented,
among others, the sample of 203 type~II Cepheids in the LMC\footnote{The
catalog of type~II Cepheids in the LMC was extended by six objects
compared to the sample published originally.} (Soszyñski \etal 2008b,
hereafter Paper~I) and the catalog of 4630 classical Cepheids in the SMC
(Soszyñski \etal 2010, hereafter Paper~II).

In Paper~I we defined a class of peculiar W~Vir stars. These objects have
distinctive light curves, are brighter and bluer than the regular W~Vir
variables. High proportion of the peculiar W~Vir stars shows evidences of
binarity (eclipsing or ellipsoidal variations). Matsunaga \etal (2009)
noticed that the bright Galactic type~II Cepheid, $\kappa$~Pav, is likely
to belong to this peculiar W~Vir class. In this paper we confirm the
classification proposed in Paper~I. In the SMC seven peculiar W~Vir stars
are cataloged.

\Section{Observational Data}
The catalog has been prepared based on the photometric data collected in
the course of the OGLE-III project with the 1.3-meter Warsaw telescope
located at Las Campanas Observatory, Chile. The observatory is operated by
the Carnegie Institution of Washington. The camera used in this stage of
the OGLE project was a mosaic consisting of eight SITe $2048\times4096$ CCD
detectors with a total field of view of about $35\times35.5$~arcmin.
Details of the instrumentation setup can be found in Udalski (2003).

The OGLE-III fields in the SMC cover about 14 square degrees. This region
was photometrically monitored from June 2001 to May 2009. Observations were
obtained in two standard bands -- {\it I} and {\it V} -- however the
sampling in the {\it I}-band is much denser (typically about 700 points)
than in the {\it V}-band (50--70 points). The data were reduced using the
Difference Image Analysis (DIA, Alard and Lupton 1998, Wo¼niak 2000). For
more details on the data reductions, photometric calibrations and
astrometric transformations see Udalski \etal (2008a).

The photometry of stars located within the central 2.4 square degrees
of the SMC was supplemented with the OGLE-II observations (Szymañski 2005)
obtained between 1997 and 2000. For individual objects the OGLE-II
photometry was adjusted to agree with the OGLE-III light curves.

\vspace*{5pt}
\Section{Selection and Classification of Type~II Cepheids}
\vspace*{5pt}
The selection process of type~II Cepheids was similar to the procedure of
identification of classical Cepheids in the SMC (Paper~II). In brief, we
performed a massive period search for over 6 million stars, \ie all objects
photometrically monitored in the SMC by the OGLE-III survey (Udalski \etal
2008b). These computations were carried out at the Interdisciplinary Centre
for Mathematical and Computational Modelling (ICM) of the University of
Warsaw. Stars with the most significant signal of periodicity
(signal-to-noise ratio larger than 5) were used to construct the
period--luminosity~(PL) diagrams in {\it V}, {\it I}, and
reddening-independent Wesenheit index $W_I=I-1.55(V-I)$. Then, the light
curves in the wide strip covering the PL relations for classical and
type~II Cepheids were visually examined and divided into pulsating-like,
eclipsing-like and other variable stars.

\begin{figure}[p]
\vspace{-0.8cm}
\centerline{\includegraphics[width=15.5cm]{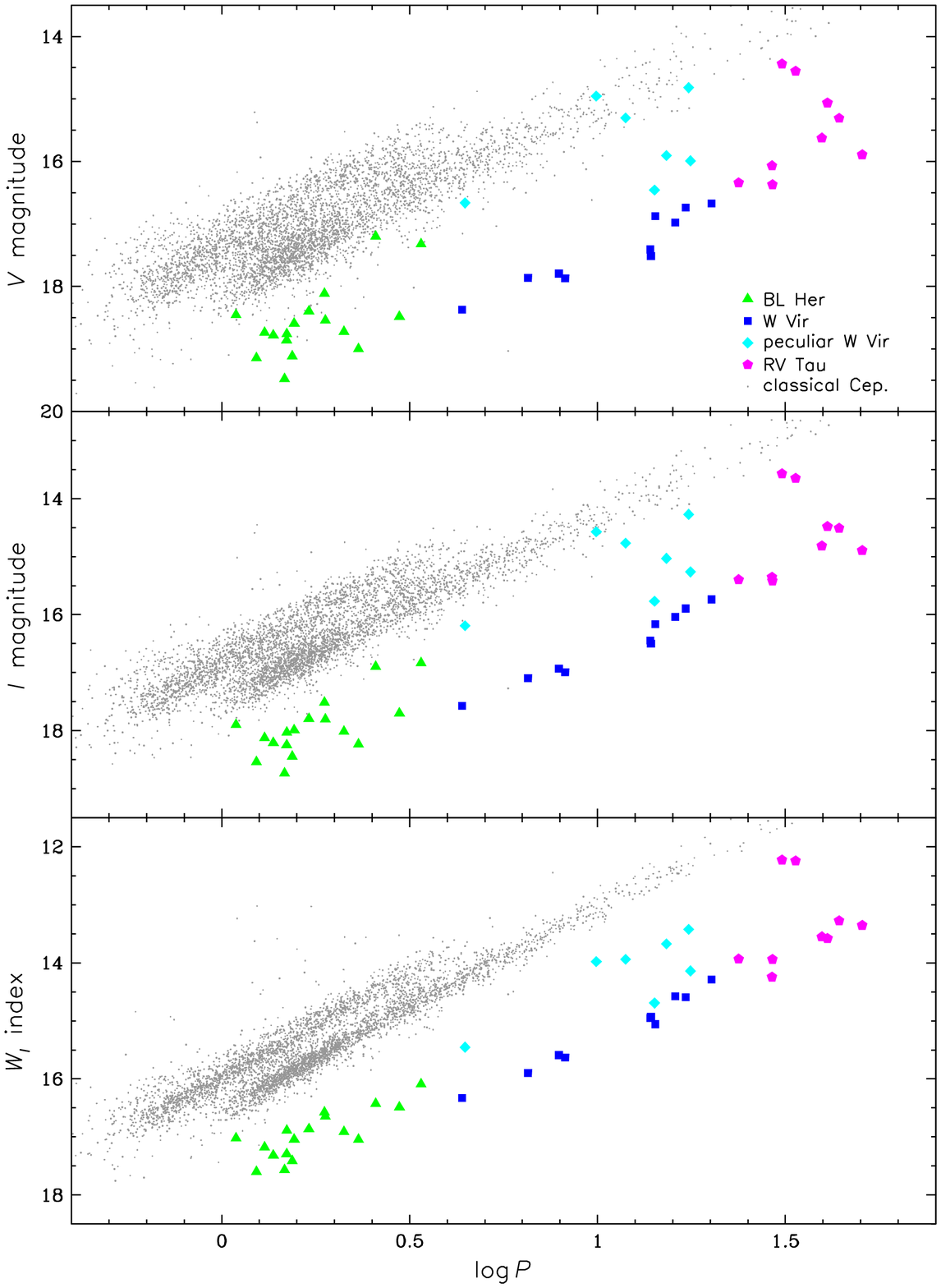}}
\vspace{-0.8cm}
\FigCap{Period--luminosity relations for Cepheids in the SMC. Green, blue
and magenta symbols represent BL~Her, W~Vir and RV~Tau stars,
respectively. Cyan diamonds indicate peculiar W~Vir stars. Small gray
points show classical Cepheids cataloged in Paper~II.}
\end{figure}

The stars classified as possible pulsating variables were examined in
detail and categorized into several groups: classical Cepheids (Paper~II),
type~II Cepheids, RR~Lyr stars, $\delta$~Sct stars and long-period
variable stars. The final classification was based on the light curve
morphology, luminosities and colors of the stars, including near-infrared
data from the IRSF/SIRIUS survey (Kato \etal 2007). Objects with
measurable proper motions were recognized as Galactic stars in the
foreground of the SMC and removed from the sample.

\begin{figure}[htb]
\centerline{\includegraphics[width=12.7cm, bb=40 260 565 745]{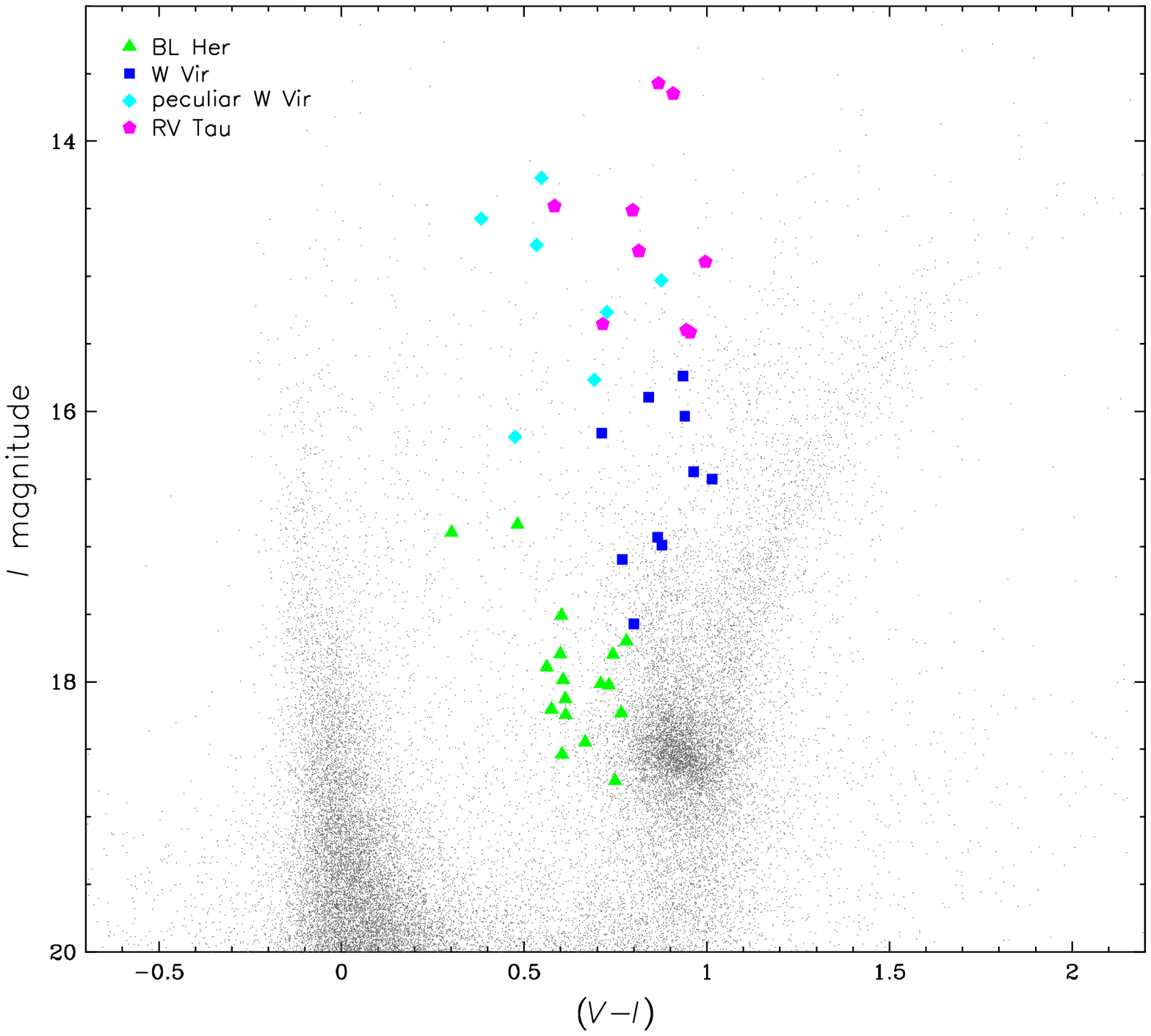}}
\FigCap{Color--magnitude diagram for type~II Cepheids in the SMC. Color
symbols represent the same type of pulsators as in Fig.~1. The background
gray points show stars from the subfield SMC100.1.}
\end{figure}

Our final list of type~II Cepheids in the SMC contains 43 objects. We
divided this sample into BL~Her, ordinary W~Vir, peculiar W~Vir and RV~Tau
stars using generally the same criteria as in the LMC catalog (Paper~I). As
a boundary between RR~Lyr and BL~Her stars we adopted a period of
1~day. The transition between BL~Her and W~Vir stars was defined at
$P=4$~days and the upper boundary for W~Vir stars was adopted at
$P=20$~days. The only exception was object OGLE-SMC-T2CEP-34 (with period
of $20.12$~days) which was classified as a W~Vir star, because its light
curve resembled that of W~Vir rather than of RV~Tau star.

\begin{figure}[p]
\vglue-0.8cm
\centerline{\includegraphics[width=15.5cm]{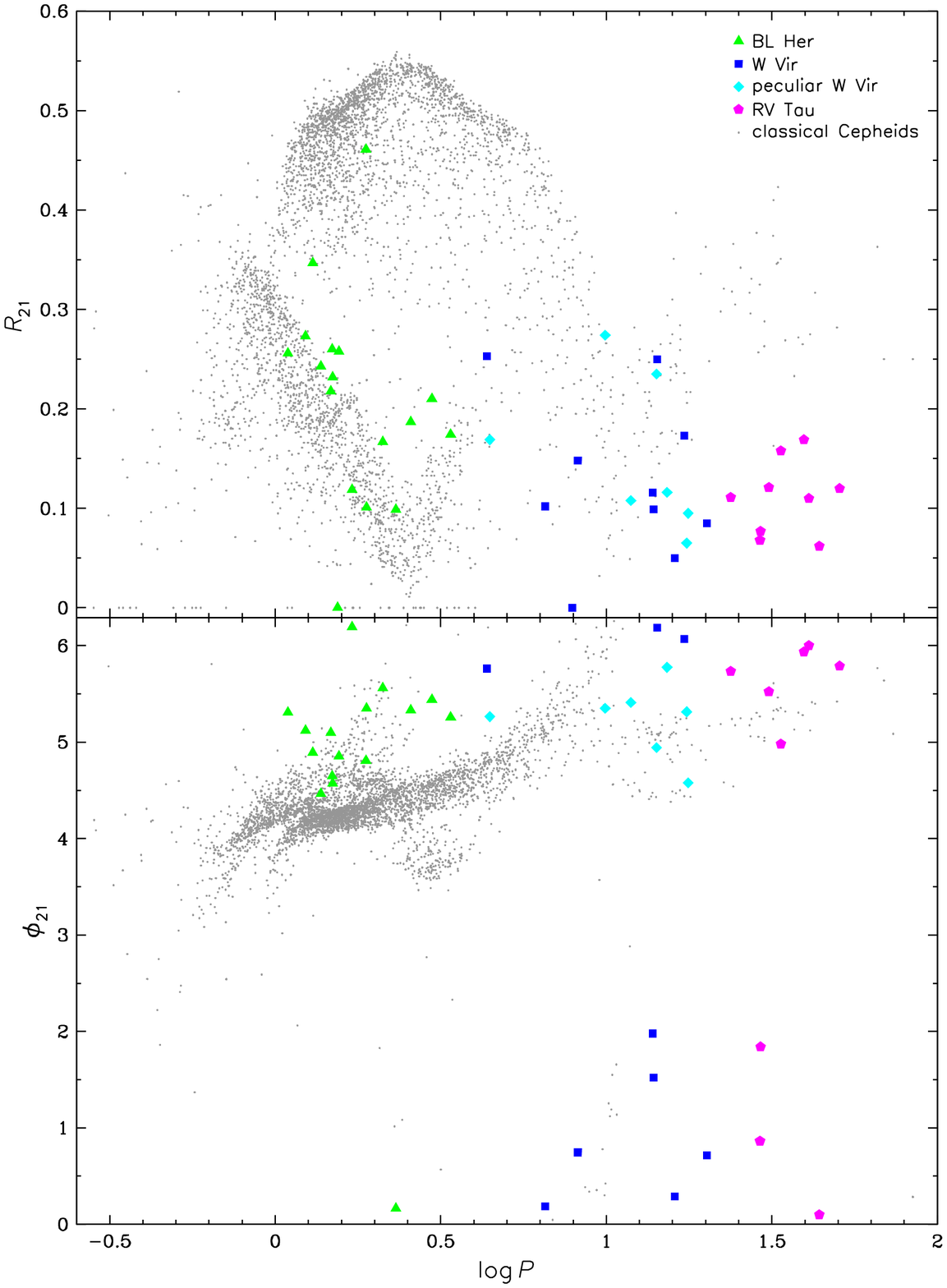}}
\vglue-0.7cm
\FigCap{Fourier parameters $R_{21}$ and $\phi_{21}$ \vs $\log{P}$ for
Cepheids in the SMC. Symbols represent the same type of stars as in
Fig.~1.}
\end{figure}

Stars with longer periods were classified in our catalog as RV~Tau
variables, although only three of them show clear alternation of deep and
shallow minima, \ie the characteristic feature of RV~Tau stars. These three
objects are the first definitive RV~Tau stars known in the SMC. Among the
other six long-period type~II Cepheids, four display semiregular light
curves and sometimes are referred to as SRd stars (yellow semiregulars).
There is no doubt that RV~Tau and SRd stars are closely related (\eg Percy
and Mohammed 2004). Both classes are represented by old, low-mass,
pulsating supergiants, populating the same range of magnitudes and colors,
and obeying the same PL relation. The remaining two stars are pulsating
variables with eclipsing or ellipsoidal modulation (see Section~6). Their
light curves, after prewhitening with the additional periodicities, are quite
stable, and do not show alternations of cycles. Hereafter, all these
long-period type~II Cepheids are called RV~Tau stars.

As in the LMC we divided W~Vir stars into two groups: regular and peculiar
variables. Objects of the latter class generally do not obey the PL
relation of type~II Cepheids (Fig.~1) -- they are brighter and bluer than
regular W~Vir variables (Fig.~2). Peculiar W~Vir stars differ from regular
variable stars in the morphology of light curves, which is especially
reflected in the Fourier coefficient $\phi_{21}$ (Simon and Lee 1981). In
the $\log{P}$--$\phi_{21}$ diagram (Fig.~3) regular and peculiar W~Vir
stars are well separated. The difference between LMC and SMC peculiar W~Vir
stars seems to be in a different range of periods where these objects appear. In
the LMC we did not find any peculiar W~Vir star with a period longer than
10~days, while in the SMC we identified objects of that kind with periods
ranging from 4.4 to 17.7~days, \ie practically in the whole range of
periods covered by W~Vir stars. Moreover, two type~II Cepheids with periods
longer than 30~days (OGLE-SMC-T2CEP-07 and OGLE-SMC-T2CEP-29), and
classified as RV~Tau variables, show features similar to the peculiar W~Vir
stars (see Section~6).

\Section{The Catalog}
The OGLE-III catalog of type~II Cepheids in the SMC comprises 43 objects:
17 BL~Her, 17 W~Vir (10 regular and 7 peculiar stars) and 9 RV~Tau
stars. Table~1 lists all the variables with their identifications in the
OGLE and GCVS databases. The stars are organized in order of increasing
right ascension and designated with symbols OGLE-SMC-T2CEP-NN, where NN is
a two-digit consecutive number. The columns of the Table~1 are: object's
designation, OGLE-III field and internal database number of the star
(consistent with the photometric maps of the SMC by Udalski \etal 2008b),
classification (BL~Her, W~Vir, peculiar W~Vir, RV~Tau), right ascension and
declination for epoch J2000.0, cross-identifications with the OGLE-II
photometric database (Szymañski 2005), cross-identifications with the GCVS
(Artyukhina \etal 1995), and other designations taken from the GCVS.

In Table~2 we present the basic observational parameters of the type~II
Cepheids from our catalog. The columns are as follows: {\it I}- and {\it
V}-band intensity mean magnitudes, ``single'' pulsation periods (\ie
intervals between successive minima) with uncertainties (derived with the
{\sc Tatry} code of Schwarzenberg-Czerny 1996), epochs of a maximum of
light, peak-to-peak amplitudes in the {\it I}-band, and Fourier parameters
$R_{21}$, $\phi_{21}$, $R_{31}$, $\phi_{31}$ (Simon and Lee 1981) derived
also for the {\it I}-band light curves.

\begin{landscape}
\renewcommand{\arraystretch}{1.15}
\MakeTable{
l@{\hspace{15pt}}
l@{\hspace{15pt}}
r@{\hspace{15pt}}
r@{\hspace{15pt}}
l@{\hspace{15pt}}
l@{\hspace{15pt}}
l@{\hspace{15pt}}
c@{\hspace{15pt}}l}
{12.5cm}{Type~II Cepheids in the SMC}
{\hline
 \multicolumn{1}{c}{Star name}
&\multicolumn{2}{c}{OGLE-III ID} 
&\multicolumn{1}{c}{Type} 
&\multicolumn{1}{c}{RA} 
&\multicolumn{1}{c}{DEC} 
&\multicolumn{1}{c}{OGLE-II ID} 
& GCVS 
& Other \\
&\multicolumn{1}{c}{Field} 
&\multicolumn{1}{c}{No.} & 
&\multicolumn{1}{c}{[J2000.0]} 
&\multicolumn{1}{c}{[J2000.0]} &
&SMC...
&designation \\
\hline
OGLE-SMC-T2CEP-01 & SMC133.6 &    32 & pWVir & 00\uph30\upm02\zdot\ups36 & $-73\arcd49\arcm07\zdot\arcs0$ &                &       & \\
OGLE-SMC-T2CEP-02 & SMC131.1 &   737 & BLHer & 00\uph34\upm53\zdot\ups51 & $-72\arcd58\arcm45\zdot\arcs9$ &                &       & \\
OGLE-SMC-T2CEP-03 & SMC133.4 &  9322 &  WVir & 00\uph37\upm08\zdot\ups35 & $-73\arcd43\arcm04\zdot\arcs9$ & SMC\_SC1\_33895  &       & \\
OGLE-SMC-T2CEP-04 & SMC125.6 &    89 &  WVir & 00\uph38\upm20\zdot\ups36 & $-73\arcd17\arcm16\zdot\arcs4$ & SMC\_SC1\_77980  &       & \\
OGLE-SMC-T2CEP-05 & SMC128.2 &   185 &  WVir & 00\uph42\upm03\zdot\ups81 & $-74\arcd01\arcm24\zdot\arcs6$ &                &       & \\
OGLE-SMC-T2CEP-06 & SMC128.4 &  2107 & BLHer & 00\uph42\upm16\zdot\ups01 & $-73\arcd39\arcm13\zdot\arcs5$ & SMC\_SC2\_81443  &       & \\
OGLE-SMC-T2CEP-07 & SMC125.4 &     2 & RVTau & 00\uph42\upm57\zdot\ups49 & $-73\arcd07\arcm14\zdot\arcs8$ & SMC\_SC3\_31999  & V0257 & HV~1369 \\
OGLE-SMC-T2CEP-08 & SMC125.2 & 35239 & BLHer & 00\uph44\upm00\zdot\ups77 & $-73\arcd22\arcm54\zdot\arcs4$ & SMC\_SC3\_130452 &       & \\
OGLE-SMC-T2CEP-09 & SMC126.1 &  1382 & BLHer & 00\uph44\upm12\zdot\ups37 & $-72\arcd59\arcm28\zdot\arcs5$ & SMC\_SC3\_157235 &       & \\
OGLE-SMC-T2CEP-10 & SMC125.2 & 13188 & pWVir & 00\uph44\upm54\zdot\ups66 & $-73\arcd28\arcm03\zdot\arcs3$ & SMC\_SC3\_184902 &       & \\
OGLE-SMC-T2CEP-11 & SMC125.1 & 37280 & pWVir & 00\uph45\upm19\zdot\ups56 & $-73\arcd30\arcm03\zdot\arcs4$ & SMC\_SC3\_184897 &       & \\
OGLE-SMC-T2CEP-12 & SMC125.2 & 50773 & RVTau & 00\uph45\upm19\zdot\ups69 & $-73\arcd20\arcm13\zdot\arcs7$ & SMC\_SC3\_193335 &       & \\
OGLE-SMC-T2CEP-13 & SMC126.1 & 19084 &  WVir & 00\uph45\upm41\zdot\ups37 & $-72\arcd58\arcm35\zdot\arcs0$ & SMC\_SC4\_32479  &       & \\
OGLE-SMC-T2CEP-14 & SMC100.8 & 45191 &  WVir & 00\uph48\upm02\zdot\ups83 & $-73\arcd21\arcm17\zdot\arcs7$ & SMC\_SC4\_159889 &       & \\
OGLE-SMC-T2CEP-15 & SMC100.7 & 58875 & BLHer & 00\uph49\upm36\zdot\ups92 & $-73\arcd10\arcm01\zdot\arcs4$ & SMC\_SC5\_111664 &       & \\
OGLE-SMC-T2CEP-16 & SMC101.1 & 32853 & BLHer & 00\uph50\upm12\zdot\ups58 & $-72\arcd43\arcm12\zdot\arcs4$ & SMC\_SC5\_235485 &       & \\
OGLE-SMC-T2CEP-17 & SMC102.4 &   258 & BLHer & 00\uph50\upm42\zdot\ups03 & $-71\arcd39\arcm18\zdot\arcs4$ &                &       & \\
OGLE-SMC-T2CEP-18 & SMC103.3 &    16 & RVTau & 00\uph51\upm07\zdot\ups23 & $-73\arcd41\arcm33\zdot\arcs4$ &                &       & \\
OGLE-SMC-T2CEP-19 & SMC103.3 & 33136 & RVTau & 00\uph53\upm27\zdot\ups69 & $-73\arcd38\arcm09\zdot\arcs5$ &                &       & \\
OGLE-SMC-T2CEP-20 & SMC101.2 & 48106 & RVTau & 00\uph53\upm35\zdot\ups98 & $-72\arcd34\arcm21\zdot\arcs8$ & SMC\_SC6\_246654 & V1017 & HV~1586 \\
OGLE-SMC-T2CEP-21 & SMC105.7 & 27734 & BLHer & 00\uph53\upm49\zdot\ups69 & $-72\arcd47\arcm37\zdot\arcs4$ & SMC\_SC6\_306714 &       & \\
\hline}
\setcounter{table}{0}
\MakeTable{
l@{\hspace{15pt}}
l@{\hspace{15pt}}
r@{\hspace{15pt}}
r@{\hspace{15pt}}
l@{\hspace{15pt}}
l@{\hspace{15pt}}
l@{\hspace{15pt}}
c@{\hspace{15pt}}l}{12.5cm}{Concluded}
{\hline
 \multicolumn{1}{c}{Star name}
&\multicolumn{2}{c}{OGLE-III ID} 
&\multicolumn{1}{c}{Type} 
&\multicolumn{1}{c}{RA} 
&\multicolumn{1}{c}{DEC} 
&\multicolumn{1}{c}{OGLE-II ID} 
& GCVS 
& Other \\
&\multicolumn{1}{c}{Field} 
&\multicolumn{1}{c}{No.} & 
&\multicolumn{1}{c}{[J2000.0]} 
&\multicolumn{1}{c}{[J2000.0]} &
&SMC...
&designation \\
\hline
OGLE-SMC-T2CEP-22 & SMC107.6 &  1433 & BLHer & 00\uph54\upm46\zdot\ups72 & $-73\arcd48\arcm32\zdot\arcs6$ &                &       & \\
OGLE-SMC-T2CEP-23 & SMC106.5 &  8159 & pWVir & 00\uph55\upm01\zdot\ups63 & $-73\arcd09\arcm47\zdot\arcs2$ & SMC\_SC7\_13470  &       & \\
OGLE-SMC-T2CEP-24 & SMC106.7 & 31830 & RVTau & 00\uph55\upm20\zdot\ups55 & $-73\arcd21\arcm37\zdot\arcs9$ & SMC\_SC7\_75158  & V1193 & HV~1647 \\
OGLE-SMC-T2CEP-25 & SMC106.6 & 34299 & pWVir & 00\uph55\upm58\zdot\ups57 & $-73\arcd12\arcm29\zdot\arcs7$ & SMC\_SC7\_83050  &       & \\
OGLE-SMC-T2CEP-26 & SMC108.7 & 50479 & BLHer & 00\uph57\upm24\zdot\ups86 & $-72\arcd13\arcm17\zdot\arcs8$ &                &       & \\
OGLE-SMC-T2CEP-27 & SMC106.8 & 38983 & BLHer & 00\uph57\upm28\zdot\ups64 & $-73\arcd31\arcm26\zdot\arcs9$ &                &       & \\
OGLE-SMC-T2CEP-28 & SMC106.8 & 38374 & pWVir & 00\uph57\upm31\zdot\ups84 & $-73\arcd32\arcm11\zdot\arcs3$ &                &       & \\
OGLE-SMC-T2CEP-29 & SMC108.2 &     3 & RVTau & 00\uph57\upm38\zdot\ups09 & $-72\arcd18\arcm12\zdot\arcs2$ & SMC\_SC8\_52799  & V1372 & HV~12140 \\
OGLE-SMC-T2CEP-30 & SMC106.5 & 51449 & BLHer & 00\uph57\upm40\zdot\ups76 & $-73\arcd03\arcm04\zdot\arcs9$ & SMC\_SC8\_3848   &       & \\
OGLE-SMC-T2CEP-31 & SMC109.4 &   174 &  WVir & 00\uph58\upm54\zdot\ups15 & $-71\arcd22\arcm56\zdot\arcs8$ &                &       & \\
OGLE-SMC-T2CEP-32 & SMC106.1 & 13852 &  WVir & 00\uph58\upm58\zdot\ups73 & $-73\arcd33\arcm45\zdot\arcs8$ &                &       & \\
OGLE-SMC-T2CEP-33 & SMC105.4 & 24131 & BLHer & 00\uph59\upm03\zdot\ups09 & $-72\arcd28\arcm32\zdot\arcs2$ & SMC\_SC8\_148923 &       & \\
OGLE-SMC-T2CEP-34 & SMC108.2 & 46300 &  WVir & 01\uph00\upm06\zdot\ups44 & $-72\arcd13\arcm51\zdot\arcs7$ & SMC\_SC8\_209984 &       & \\
OGLE-SMC-T2CEP-35 & SMC107.3 &  8878 &  WVir & 01\uph00\upm35\zdot\ups01 & $-73\arcd46\arcm57\zdot\arcs9$ &                & V1577 & HV~1828 \\
OGLE-SMC-T2CEP-36 & SMC111.5 &   288 & BLHer & 01\uph02\upm40\zdot\ups96 & $-73\arcd10\arcm02\zdot\arcs6$ &                &       & \\
OGLE-SMC-T2CEP-37 & SMC112.8 &   567 & BLHer & 01\uph03\upm46\zdot\ups50 & $-74\arcd07\arcm28\zdot\arcs8$ &                &       & \\
OGLE-SMC-T2CEP-38 & SMC111.8 &  5979 & pWVir & 01\uph04\upm29\zdot\ups00 & $-73\arcd33\arcm53\zdot\arcs6$ &                &       & \\
OGLE-SMC-T2CEP-39 & SMC111.4 &   165 & BLHer & 01\uph06\upm40\zdot\ups91 & $-73\arcd07\arcm05\zdot\arcs0$ & SMC\_SC11\_100   &       & \\
OGLE-SMC-T2CEP-40 & SMC119.8 &   119 &  WVir & 01\uph08\upm46\zdot\ups85 & $-71\arcd51\arcm09\zdot\arcs4$ &                &       & \\
OGLE-SMC-T2CEP-41 & SMC116.6 & 11058 & RVTau & 01\uph13\upm19\zdot\ups05 & $-73\arcd18\arcm07\zdot\arcs8$ &                &       & \\
OGLE-SMC-T2CEP-42 & SMC123.3 &  2298 & BLHer & 01\uph23\upm26\zdot\ups69 & $-72\arcd00\arcm24\zdot\arcs3$ &                &       & \\
OGLE-SMC-T2CEP-43 & SMC123.2 &  2021 & RVTau & 01\uph23\upm53\zdot\ups06 & $-72\arcd16\arcm45\zdot\arcs9$ &                &       & \\
\hline}
\end{landscape}
\noindent

\renewcommand{\TableFont}{\scriptsize}
\renewcommand{\arraystretch}{1.12}
\MakeTable{l@{\hspace{4pt}}
c@{\hspace{4pt}}
c@{\hspace{3pt}}
r@{\hspace{3pt}}
r@{\hspace{0pt}}
r@{\hspace{4pt}}
c@{\hspace{4pt}}
c@{\hspace{3pt}}
c@{\hspace{4pt}}
c@{\hspace{3pt}}
c}{12.5cm}{Observational parameters of type~II Cepheids in the SMC}
{\hline
\noalign{\vskip3pt}
\multicolumn{1}{c}{Star name} & $\langle{I}\rangle$ & $\langle{V}\rangle$ &
\multicolumn{1}{c}{$P$} & \multicolumn{1}{c}{$\sigma_P$} &
\multicolumn{1}{c}{$T_{\rm max}$} & $A_I$ & $R_{21}$ & $\phi_{21}$ & $R_{31}$ & $\phi_{31}$ \\
& [mag] & [mag] & \multicolumn{1}{c}{[days]} & \multicolumn{1}{c}{[days]} & \multicolumn{1}{c}{\tiny HJD-2450000} & [mag] & & & & \\
\noalign{\vskip3pt}
\hline
\noalign{\vskip3pt}
OGLE-SMC-T2CEP-01 & 14.768 & 15.303 & 11.8690178 & 0.0003296 & 2088.65297 & 0.191 & 0.108 & 5.411 & 0.018 & 5.826 \\
OGLE-SMC-T2CEP-02 & 18.206 & 18.781 &  1.3721870 & 0.0000030 & 2102.59657 & 0.514 & 0.243 & 4.466 & 0.074 & 5.697 \\
OGLE-SMC-T2CEP-03 & 17.572 & 18.373 &  4.3598789 & 0.0000114 &  627.46722 & 0.407 & 0.253 & 5.763 & 0.104 & 5.264 \\
OGLE-SMC-T2CEP-04 & 17.093 & 17.862 &  6.5333997 & 0.0001285 &  624.25665 & 0.077 & 0.102 & 0.185 & 0.080 & 6.253 \\
OGLE-SMC-T2CEP-05 & 16.990 & 17.867 &  8.2058890 & 0.0001764 & 2087.54105 & 0.140 & 0.148 & 0.746 & 0.036 & 0.596 \\
OGLE-SMC-T2CEP-06 & 18.536 & 19.140 &  1.2356136 & 0.0000013 &  625.69012 & 0.523 & 0.273 & 5.123 & 0.150 & 3.970 \\
OGLE-SMC-T2CEP-07 & 13.572 & 14.440 & 30.9606438 & 0.0004026 &  596.44583 & 0.415 & 0.121 & 5.524 & 0.020 & 3.670 \\
OGLE-SMC-T2CEP-08 & 18.025 & 18.757 &  1.4897859 & 0.0000038 &  624.55707 & 0.206 & 0.232 & 4.572 & 0.050 & 2.952 \\
OGLE-SMC-T2CEP-09 & 17.700 & 18.480 &  2.9710719 & 0.0000057 &  618.99334 & 0.356 & 0.210 & 5.440 & 0.123 & 4.397 \\
OGLE-SMC-T2CEP-10 & 14.272 & 14.820 & 17.4807419 & 0.0005562 &  619.79231 & 0.054 & 0.065 & 5.315 & 0.000 & \multicolumn{1}{c}{--} \\
OGLE-SMC-T2CEP-11 & 14.571 & 14.954 &  9.9253977 & 0.0000552 &  617.95658 & 0.197 & 0.274 & 5.352 & 0.153 & 4.325 \\
OGLE-SMC-T2CEP-12 & 15.415 & 16.369 & 29.2189113 & 0.0039443 &  608.57963 & 0.103 & 0.077 & 1.840 & 0.022 & 4.392 \\
OGLE-SMC-T2CEP-13 & 16.444 & 17.408 & 13.8099916 & 0.0001602 &  617.96318 & 0.489 & 0.116 & 1.978 & 0.079 & 2.634 \\
OGLE-SMC-T2CEP-14 & 16.501 & 17.516 & 13.8783799 & 0.0003583 &  622.23967 & 0.456 & 0.099 & 1.521 & 0.073 & 2.624 \\
OGLE-SMC-T2CEP-15 & 16.895 & 17.196 &  2.5695964 & 0.0000054 &  464.73357 & 0.172 & 0.187 & 5.333 & 0.080 & 3.686 \\
OGLE-SMC-T2CEP-16 & 18.013 & 18.723 &  2.1131980 & 0.0000020 &  464.88124 & 0.461 & 0.167 & 5.564 & 0.076 & 3.589 \\
OGLE-SMC-T2CEP-17 & 18.126 & 18.739 &  1.2993097 & 0.0000099 & 2540.43032 & 0.320 & 0.347 & 4.893 & 0.074 & 3.563 \\
OGLE-SMC-T2CEP-18 & 14.813 & 15.627 & 39.5193655 & 0.0057005 & 2048.45132 & 0.667 & 0.169 & 5.936 & 0.073 & 5.861 \\
OGLE-SMC-T2CEP-19 & 14.481 & 15.064 & 40.9118097 & 0.0059419 & 2055.86074 & 0.136 & 0.110 & 6.003 & 0.048 & 6.017 \\
OGLE-SMC-T2CEP-20 & 14.894 & 15.890 & 50.6231456 & 0.0069705 &  439.44469 & 0.750 & 0.120 & 5.789 & 0.032 & 5.801 \\
OGLE-SMC-T2CEP-21 & 18.233 & 18.999 &  2.3131264 & 0.0000203 &  465.02486 & 0.097 & 0.099 & 0.164 & 0.000 & \multicolumn{1}{c}{--} \\
OGLE-SMC-T2CEP-22 & 18.731 & 19.480 &  1.4705261 & 0.0000067 & 2085.76244 & 0.410 & 0.218 & 5.098 & 0.112 & 0.590 \\
OGLE-SMC-T2CEP-23 & 15.264 & 15.990 & 17.6752925 & 0.0005561 &  604.25442 & 0.121 & 0.095 & 4.576 & 0.000 & \multicolumn{1}{c}{--} \\
OGLE-SMC-T2CEP-24 & 14.511 & 15.308 & 43.9607975 & 0.0179727 &  615.47093 & 0.154 & 0.062 & 0.101 & 0.127 & 3.929 \\
OGLE-SMC-T2CEP-25 & 15.766 & 16.458 & 14.1708916 & 0.0001848 &  607.92509 & 0.481 & 0.235 & 4.944 & 0.065 & 4.436 \\
OGLE-SMC-T2CEP-26 & 17.794 & 18.393 &  1.7048368 & 0.0000117 & 2085.92662 & 0.133 & 0.119 & 6.194 & 0.000 & \multicolumn{1}{c}{--} \\
OGLE-SMC-T2CEP-27 & 18.446 & 19.113 &  1.5417249 & 0.0000080 & 2086.43490 & 0.291 & 0.000 & \multicolumn{1}{c}{--} & 0.160 & 1.387 \\
OGLE-SMC-T2CEP-28 & 15.031 & 15.907 & 15.2643255 & 0.0003658 & 2093.82590 & 0.194 & 0.116 & 5.775 & 0.016 & 4.732 \\
OGLE-SMC-T2CEP-29 & 13.648 & 14.556 & 33.6764628 & 0.0004650 &  600.72415 & 0.492 & 0.168 & 4.982 & 0.072 & 3.447 \\
OGLE-SMC-T2CEP-30 & 16.836 & 17.318 &  3.3889388 & 0.0000064 &  620.05824 & 0.245 & 0.174 & 5.259 & 0.105 & 3.588 \\
OGLE-SMC-T2CEP-31 & 16.929 & 17.794 &  7.8953026 & 0.0004850 & 2079.41753 & 0.039 & 0.000 & \multicolumn{1}{c}{--} & 0.000 & \multicolumn{1}{c}{--} \\
OGLE-SMC-T2CEP-32 & 16.161 & 16.873 & 14.2468347 & 0.0005521 & 2073.05395 & 0.604 & 0.250 & 6.188 & 0.064 & 5.056 \\
OGLE-SMC-T2CEP-33 & 17.510 & 18.112 &  1.8776865 & 0.0000014 &  620.56334 & 0.508 & 0.461 & 4.809 & 0.214 & 3.278 \\
OGLE-SMC-T2CEP-34 & 15.737 & 16.672 & 20.1206110 & 0.0003603 &  618.98689 & 0.792 & 0.085 & 0.715 & 0.040 & 2.486 \\
OGLE-SMC-T2CEP-35 & 15.894 & 16.735 & 17.1814841 & 0.0001830 & 2072.74556 & 0.909 & 0.173 & 6.069 & 0.071 & 4.216 \\
OGLE-SMC-T2CEP-36 & 17.892 & 18.454 &  1.0916592 & 0.0000017 & 2104.57149 & 0.358 & 0.256 & 5.311 & 0.065 & 4.735 \\
OGLE-SMC-T2CEP-37 & 17.986 & 18.593 &  1.5590709 & 0.0000030 & 2103.84900 & 0.496 & 0.258 & 4.856 & 0.088 & 1.029 \\
OGLE-SMC-T2CEP-38 & 16.189 & 16.664 &  4.4440180 & 0.0000152 & 2101.74437 & 0.237 & 0.169 & 5.264 & 0.113 & 3.573 \\
OGLE-SMC-T2CEP-39 & 17.798 & 18.541 &  1.8875529 & 0.0000079 &  625.40072 & 0.128 & 0.101 & 5.350 & 0.000 & \multicolumn{1}{c}{--} \\
OGLE-SMC-T2CEP-40 & 16.036 & 16.976 & 16.1110373 & 0.0002492 & 2092.07312 & 0.873 & 0.050 & 0.291 & 0.053 & 3.043 \\
OGLE-SMC-T2CEP-41 & 15.353 & 16.068 & 29.1184308 & 0.0069526 & 2089.43875 & 0.152 & 0.068 & 0.864 & 0.105 & 5.159 \\
OGLE-SMC-T2CEP-42 & 18.246 & 18.860 &  1.4874289 & 0.0000032 & 2090.49055 & 0.562 & 0.260 & 4.646 & 0.101 & 0.284 \\
OGLE-SMC-T2CEP-43 & 15.398 & 16.342 & 23.7429305 & 0.0010767 & 2068.62766 & 0.890 & 0.111 & 5.736 & 0.030 & 4.439 \\
\noalign{\vskip3pt}
\hline}

\begin{figure}[p]  
\centerline{\includegraphics[width=13.5cm]{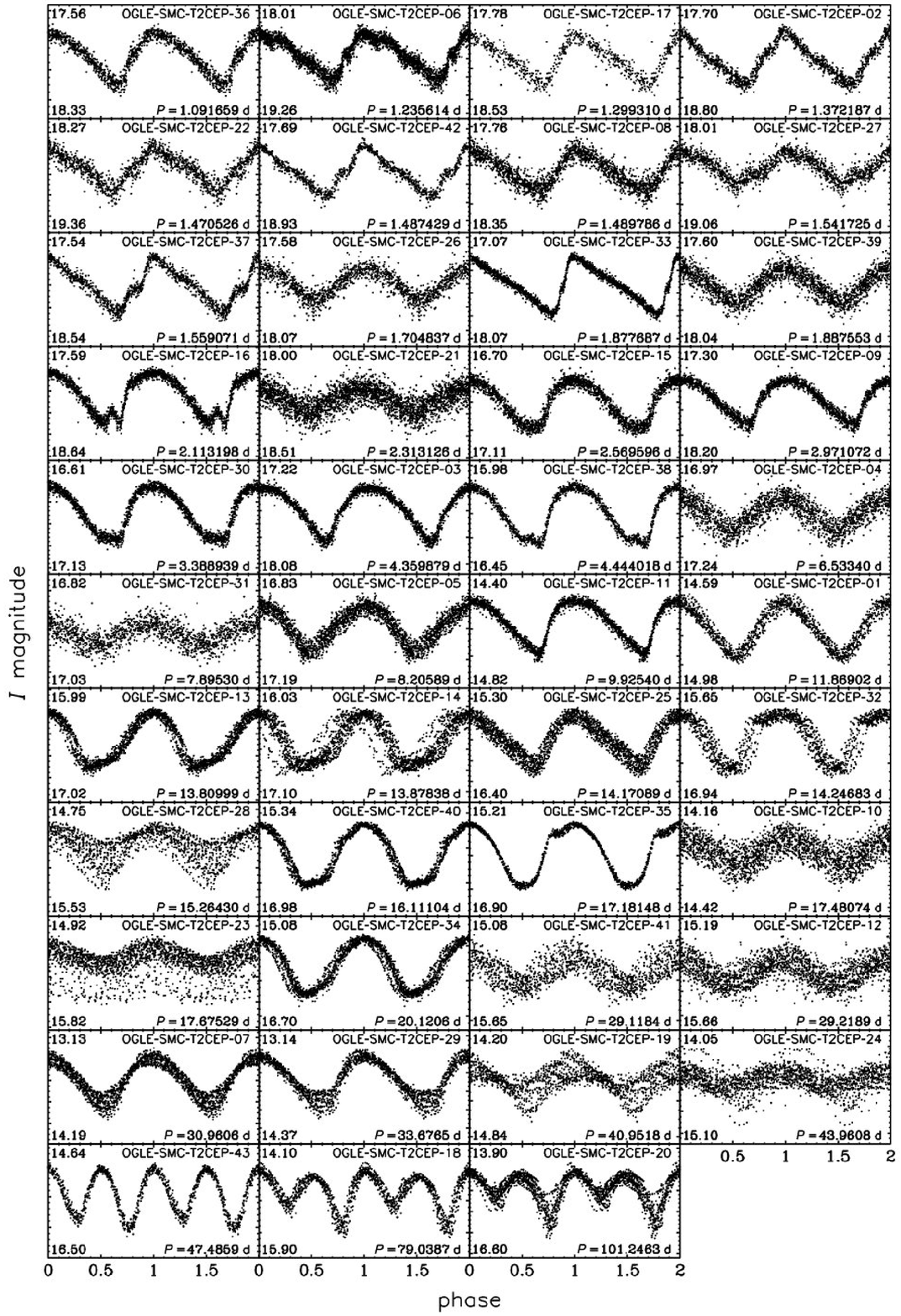}}
\FigCap{Phased {\it I}-band light curves of type~II Cepheids in the
SMC. The stars are arranged with increasing periods. The range of
magnitudes varies from panel to panel. Numbers in the left corners show the
lower and the upper limits of magnitudes.}
\end{figure}

The content of Tables 1 and 2 can be downloaded in the electronic form
through the WWW interface or by the anonymous FTP site:
\begin{center}
{\it http://ogle.astrouw.edu.pl/} \\
{\it ftp://ftp.astrouw.edu.pl/ogle/ogle3/OIII-CVS/smc/t2cep/}\\
\end{center}

The OGLE-II and OGLE-III {\it I}- and {\it V}-band photometry and finding
charts for all stars can be downloaded from the same sites.

Phased {\it I}-band light curves for all the stars from our catalog are
shown in Fig.~4. The stars are arranged with increasing periods to show the
progression of the light curve shapes with period. The data are folded with
the ``single'' periods with the exception of three RV~Tau stars that show
alternating deep and shallow minima and are presented with their formal
(``double'') periods. For multiperiodic stars the light curves are phased
with their primary pulsation periods.

\Section{Cross-Identification with Other Catalogs}
We cross-matched our sample of type~II Cepheids in the SMC with the
previously published lists of variable stars of this type. In the GCVS
Vol.~V (Artyukhina \etal 1995) five SMC stars were designated as CWA
or CWB variables, \ie type~II Cepheids. With the OGLE data we can confirm
the classification of only one object from this list (OGLE-SMC-T2CEP-35 =
HV~1828). Two of the GCVS CW stars (XY~Hyi and HV~12901) are located outside
the OGLE-III fields and no photometric information is available for them in our
database. The two remaining stars classified in the GCVS as type~II
Cepheids (WS12, WS45) turned out to be classical Cepheids and were
cataloged in the previous part of the OIII-CVS (Paper~II). Our sample also
contains several objects which had entries in the GCVS, but were not
classified as type~II Cepheids.

In the OGLE-II catalog of Cepheids in the SMC (Udalski \etal 1999) 16 stars
were classified as ``fainter'' (FA) than fundamental-mode classical
Cepheids. Most of these stars should be type~II Cepheids, although the
OGLE-II sample was selected based on only two seasons of observations, so
the classification was rather uncertain. Our catalog includes 11 objects
from the list provided by Udalski \etal (1999). The other five objects
turned out to be other kinds of periodic stars, mainly ellipsoidal
variables.

\Section{Discussion}
In Fig.~1 we present the PL diagrams in {\it V}, {\it I} and $W_I$ index
for type~II Cepheids found in the SMC. For comparison, classical Cepheids
(Paper~II) are plotted with small dots in the same diagrams. When
neglecting peculiar W~Vir stars and two RV~Tau stars residing in the
eclipsing systems, the scatter of the relation is rather small, especially
for longer period type~II Cepheids (W~Vir and RV~Tau stars). Similarly to
the LMC (Paper~I, Matsunaga \etal 2009), the PL relations are not linear
within the entire range of periods. RV~Tau stars seem to be brighter than
expected from the linear extrapolation of the PL relation for the
shorter-period type~II Cepheids. It is worth noticing that such a
non-linearity is not visible for type~II Cepheids in the Galactic globular
clusters (Matsunaga \etal 2006).

In Paper~I we distinguished a new subtype of type~II Cepheids -- peculiar
W~Vir stars. From 17 stars of this kind in the LMC four objects exhibited
additional eclipsing variations superimposed on the pulsation light
curve. Four other peculiar W~Vir stars in the LMC showed some
small-amplitude long-period modulation which might be explained by the
ellipsoidal variations due to binarity. If we take into account the fact
that we see the binarity effects only in a fraction of systems due to
random distribution of orbital inclinations, it is justified to conclude
that all peculiar W~Vir stars are likely to be members of binary systems.

The same conclusion can be drawn from the SMC sample presented in this
study. Seven variable stars in our catalog have been classified as peculiar
W~Vir stars, of which two exhibit eclipsing variations and two others show
ellipsoidal modulation. Additionally we identified two RV~Tau stars with
some signs of binarity -- one with eclipsing and another one with
ellipsoidal modulation. Tables~3 and~4 give the lists of type~II Cepheids
in eclipsing and ellipsoidal binary systems, respectively. The prewhitened
and detrended light curves of these objects are shown in Figs.~5
and~6. Since the main pulsation period of OGLE-SMC-T2CEP-25 varied during
the OGLE observations, we adjusted the periods separately in each observing
season and we prewhitened the light curve independently in each
chunk. Among the stars in Table~4, OGLE-SMC-T2CEP-07 and OGLE-SMC-T2CEP-25
are rather certainly the ellipsoidal variables, as their long-period
components have minima alternating their depths, which is a characteristic
feature of the ellipsoidal variability. For OGLE-SMC-T2CEP-10 the
longest-period variability should be treated as a candidate for ellipsoidal
modulation.

\tabcolsep=3pt
\MakeTableee{l@{\hspace{10pt}}cc@{\hspace{10pt}}r@{.}lc@{\hspace{10pt}}r@{.}lcc}{12.5cm}{Type~II Cepheids in Eclipsing Binary Systems.}
{\hline
\noalign{\vskip3pt}
\multicolumn{1}{c}{Star name} & $P_p$ & $A_I^p$ & \multicolumn{2}{c}{$P_e$} & $A_I^e$ & \multicolumn{2}{c}{$P_s$} & $A_I^s$ & $1/P_s=$ \\
 & [days] & [mag] & \multicolumn{2}{c}{[days]} & [mag] & \multicolumn{2}{c}{[days]} & [mag] & \\
\noalign{\vskip3pt}
\hline
\noalign{\vskip3pt}
OGLE-SMC-T2CEP-23 & 17.6753 & 0.121 & 156&884 & 0.392 & 22&8199 & 0.030 & $1/P_p-2/P_e$ \\
                  & & & \multicolumn{2}{c}{~} &       & 15&8846 & 0.020 & $1/P_p+1/P_e$ \\
OGLE-SMC-T2CEP-28 & 15.2643 & 0.194 & 141&835 & 0.202 & 19&4538 & 0.110 & $1/P_p-2/P_e$ \\
                  & & & \multicolumn{2}{c}{~} &       & 12&5637 & 0.022 & $1/P_p+2/P_e$ \\
                  & & & \multicolumn{2}{c}{~} &       & 17&1069 & 0.020 & $1/P_p-1/P_e$ \\
                  & & & \multicolumn{2}{c}{~} &       &  8&5526 & 0.016 & $2/P_p-2/P_e$ \\
OGLE-SMC-T2CEP-29 & 33.6765 & 0.492 & 608&6   & 0.122 & 37&8939 & 0.075 & $1/P_p-2/P_e$ \\
                  & & & \multicolumn{2}{c}{~} &       & 30&2961 & 0.068 & $1/P_p+2/P_e$ \\
                  & & & \multicolumn{2}{c}{~} &       & 33&9861 & 0.049 &  \\
                  & & & \multicolumn{2}{c}{~} &       & 33&3897 & 0.037 &  \\
                  & & & \multicolumn{2}{c}{~} &       & 15&9511 & 0.028 & $2/P_p+2/P_e$ \\
\noalign{\vskip3pt}
\hline}
\begin{figure}[b]
\centerline{\includegraphics[width=13cm]{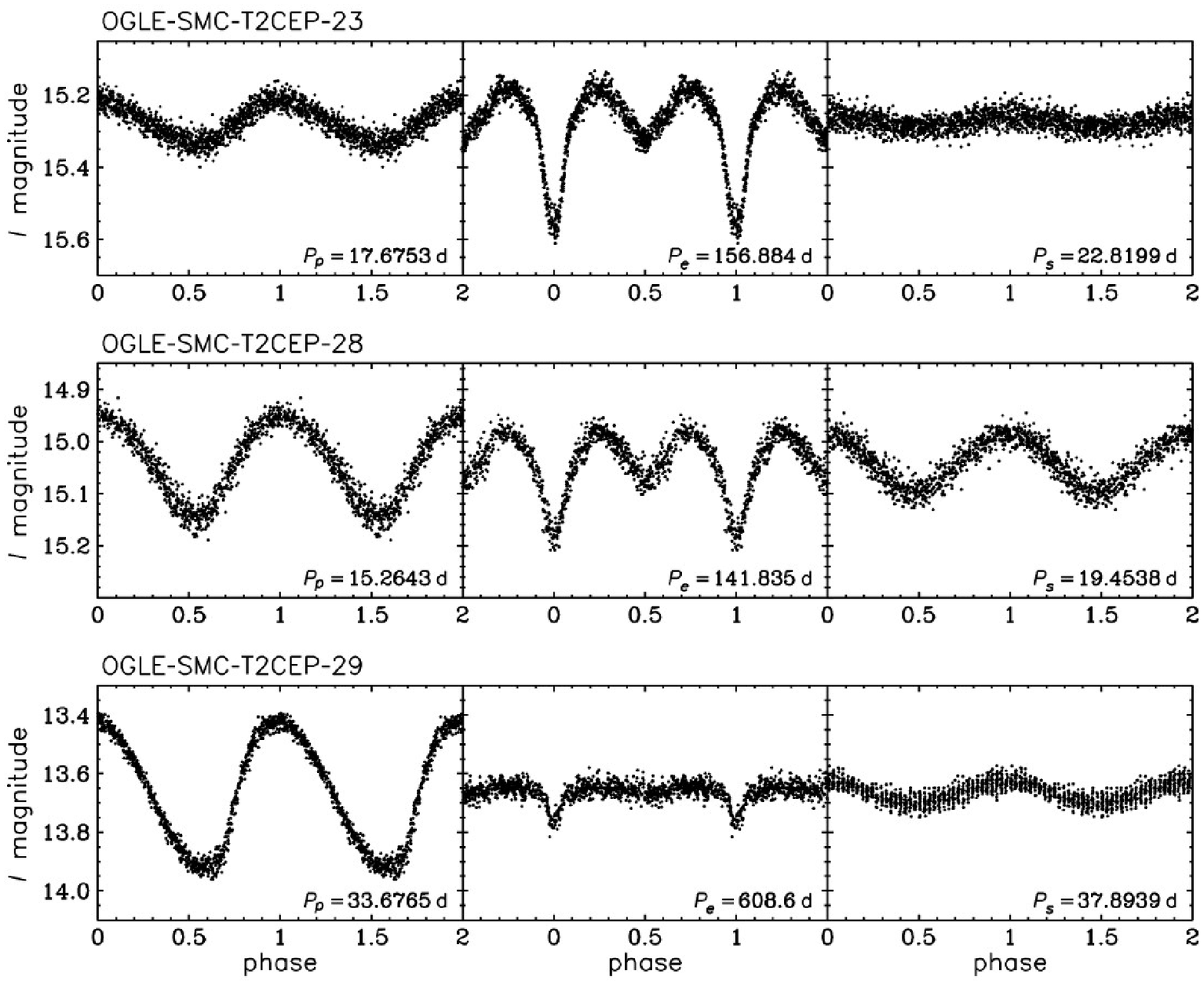}}
\FigCap{Light curves of type~II Cepheids in eclipsing binary systems. {\it Left panels} show the main pulsation light curves. {\it Middle panels} show the eclipsing light curves. {\it Right panels} show the highest-amplitude secondary components. Each light curve has been detrended and prewhitenned with all other detected periods.}
\end{figure}

\tabcolsep=3pt
\MakeTable{l@{\hspace{10pt}}r@{.}lc@{\hspace{10pt}}cc@{\hspace{10pt}}r@{.}lcc}{12.5cm}{Type~II Cepheids in Ellipsoidal Binary Systems.}
{\hline
\noalign{\vskip3pt}
\multicolumn{1}{c}{Star name} & \multicolumn{2}{c}{$P_p$} & $A_I^p$ & $P_e$ & $A_I^e$ & \multicolumn{2}{c}{$P_s$} & $A_I^s$ & $1/P_s=$ \\
 & \multicolumn{2}{c}{[days]} & [mag] & [days] & [mag] & \multicolumn{2}{c}{[days]} & [mag] & \\
\noalign{\vskip3pt}
\hline
\noalign{\vskip3pt}
OGLE-SMC-T2CEP-07 & 30&9606 & 0.415 & 392.93  & 0.068 & 36&7372 & 0.090 & $1/P_p-2/P_e$ \\
                  & \multicolumn{2}{c}{~} & & &       & 26&7370 & 0.034 & $1/P_p+2/P_e$ \\
                  & \multicolumn{2}{c}{~} & & &       & 31&1276 & 0.029 & \\
                  & \multicolumn{2}{c}{~} & & &       & 16&7994 & 0.024 & $2/P_p-2/P_e$ \\
OGLE-SMC-T2CEP-10 & 17&4807 & 0.054 & 198.18  & 0.020 & 15&7658 & 0.054 & \\
                  & \multicolumn{2}{c}{~} & & &       & 21&2285 & 0.010 & $1/P_p-2/P_e$ \\
OGLE-SMC-T2CEP-25 & 14&17089 & 0.481 & 174.87 & 0.094 & 16&9125 & 0.056 & $1/P_p-2/P_e$ \\
                  & \multicolumn{2}{c}{~} & & &       &  7&7105 & 0.014 & $2/P_p-2/P_e$ \\
                  & \multicolumn{2}{c}{~} & & &       & 12&1937 & 0.012 & $1/P_p+2/P_e$ \\
\noalign{\vskip3pt}
\hline}

\begin{figure}[b]
\centerline{\includegraphics[width=13cm]{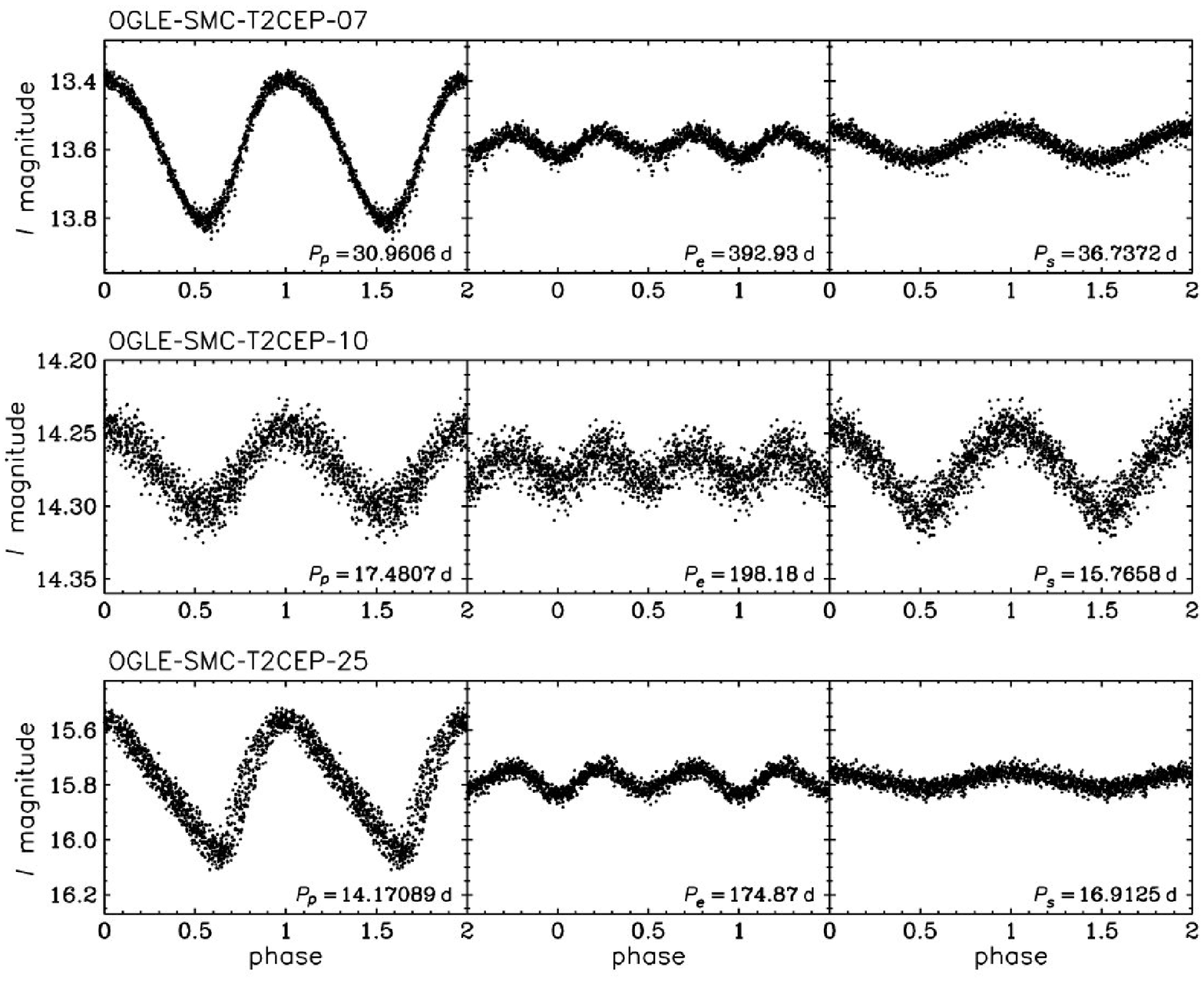}}
\FigCap{Light curves of type~II Cepheids with additional ellipsoidal
variability. {\it Left panels} show the main pulsation light curves. {\it
Middle panels} show the ellipsoidal light curves. {\it Right panels} show
the highest-amplitude secondary components. Each light curve has been
detrended and prewhitenned with all other detected periods.}
\end{figure}

The most striking feature of the type~II Cepheids in binary systems is that
all of them exhibit multiperiodicity. In Tables~3 and~4, as well as in
Figs.~5 and~6, $P_p$ indicates the primary pulsation period, $P_e$ is an
orbital period taken from eclipsing or ellipsoidal variability, while $P_s$
are the largest-amplitude secondary periods. Right columns of Figs.~5 and~6
show the light curves folded with only one, the strongest secondary
period. For all stars but OGLE-SMC-T2CEP-10 the most significant secondary
period satisfies the equation:
$$\frac{1}{P_s}=\frac{1}{P_p}-\frac{2}{P_e}.\eqno(1)$$
Other secondary periods are usually also a combination of the pulsation and
orbital period, as given in the last column of Tables~3 and~4. In the power
spectra we usually observe equidistant triplets with the main pulsational
frequency at the center and two secondary peaks on both sides at the
distance of $2/P_e$. Generally, these two peaks have different heights.
Sometimes the secondary frequencies exist at $1/P_e$ from the pulsation
peak. Power spectra of this kind may correspond to the modulation of
amplitudes and/or phases of the pulsational light curves with periods equal
to the orbital or half the orbital period.

Explanation of such behavior is beyond the scope of this paper. Detailed
analysis of these cases is necessary to enhance our knowledge about the
oscillations of stars which are distorted by tidal interactions from their
companions. Nevertheless, we can draw one important conclusion: our
Cepheids with eclipsing and ellipsoidal variations are not the optical
blends of a pulsating star with a binary system, but we discovered binaries
with a pulsating star as one of its components. In the case of
OGLE-SMC-T2CEP-10 two main periods have almost the same amplitudes and do
not obey Eq.~(1). However, we found another small-amplitude secondary
period that fulfills this rule.

The two RV~Tau stars with eclipsing or ellipsoidal modulation
(OGLE-SMC-T2CEP-07, OGLE-SMC-T2CEP-29) are rather untypical representatives
of this type of pulsating stars. Their pulsation light curves do not show
any alternations of the minima. After removing other periodicities the
light curves are quite stable (contrary to a typical RV~Tau star) and have
shapes similar to the shorter-period peculiar W~Vir stars. They are also
significantly brighter than other RV~Tau stars. We suggest that these two
objects are the equivalents of the peculiar W~Vir stars in the range of
periods occupied by RV~Tau stars.
\vspace*{-7pt}
\Section{Summary}
\vspace*{-5pt}
We performed the first search dedicated exclusively to type~II Cepheids in
the SMC and found 43 objects of this type. Thus, we increased the number of
known type~II Cepheids in the SMC by several times. We also discovered
first definitive RV~Tau stars in this galaxy.

We found three type~II Cepheids in eclipsing binary systems and three
pulsators with additional ellipsoidal variations. Note that among two
orders of magnitude more numerous sample of classical Cepheids in the SMC
(Paper~II) we identified only two pulsators with superimposed eclipsing
variability. Moreover, while among the classical Cepheids it cannot be
excluded that what we observe is the optical blend of a Cepheid and an
eclipsing variable, in the case of type~II Cepheids we are convinced that
pulsating stars are actually the members of binary systems. It is evident
because the amplitudes and/or phases of pulsation light curves are
modulated with orbital periods or their harmonics. Such a modulation must
be related to the tidal interactions between components of the binary
systems, but it requires detailed studies and modeling of this phenomenon.

The huge catalogs of classical Cepheids in the LMC (Soszyñski \etal 2008)
and the SMC (Paper~II) with the samples of type~II Cepheids in both Clouds
presented in Paper~I and in this paper give an opportunity to study in
detail these important distance indicators. Do both populations give
consistent distance scales? This is a very important task for future
investigations.

\Acknow{We would like to thank W.~Dziembowski and Z.~Ko³acz\-kowski for
discussions which clarified some issues concerning secondary periods in
Cepheids in binary systems. We are grateful to Z.~Ko³aczkowski,
A.~Schwar\-zenberg-Czerny and J.~Skowron for providing codes which enabled
us to prepare this study.

This work has been supported by MNiSW grant NN203293533. The research
leading to these results has received funding from the European Research
Council under the European Community's Seventh Framework Programme
(FP7/2007-2013)/ERC grant agreement no. 246678. The massive period
search was performed at the Interdisciplinary Centre for Mathematical and
Computational Modeling of Warsaw University (ICM), project no.~G32-3. We
wish to thank M. Cytowski for his skilled support.}

\end{document}